# Advanced eco-friendly formulations of guar biopolymer-based textile conditioners

Evdokia K. Oikonomou* and Jean-François Berret

*Université de Paris CNRS Matière et systèmes complexes 75013 Paris France*

**Abstract:** Fabric conditioners are household products used to impart softness and fragrance to textiles. They are colloidal dispersions of cationic double chain surfactants that self-assemble in vesicles. These surfactants are primarily derived from palm oil chemical modification. Reducing the content of these surfactants allows to obtain products with lower environmental impact. Such a reduction, without adverse effects on the characteristics of the softener and its performance, can be achieved by adding hydrophilic biopolymers. Here, we review the role of guar biopolymers modified with cationic or hydroxyl-propyl groups on the physicochemical properties of the formulation. Electronic and optical microscopy, dynamic light scattering, X-ray scattering and rheology of vesicles dispersion in absence and in presence of guar biopolymers are analyzed. Finally, the deposition of the new formulation on cotton fabrics is examined through scanning electron microscopy and a new protocol based on fluorescent microscopy. With this methodology it is possible to quantify the deposition of surfactants on cotton fibers. The results show that the approach followed here can facilitate the design of sustainable home-care products.

## 1. Introduction

Household products are an important class of materials for everyday life. Although surfactants, whether anionic, cationic or non-ionic are the principal ingredients of these formulations, polymers play a major role in their development. The addition of hydrophilic polymers in small amounts can improve the release of soils, inhibit the transfer of dyes and control the rheological properties, while often limiting the environmental impact of the product [1]. Polycarboxylates and polyacrylates represent an important category for dishwashing, floor and fabric cleaning products [2]. For instance, laundry detergents have been based on polycarboxylates such as polyacrylic acid homopolymer or copolymers in combination with maleic acid [3] as an environmental alternative to phosphates [1]. When it comes to fabric care products, in addition to ecological concerns, they must also be safe for humans. Fabric softeners are a category of fabric care products whose consumption is growing significantly each year on a global scale.

Fabric softeners are colloidal dispersions that give textiles softness, smoothness, flexibility, anti-static properties, water repellency and sewability. There are two main types of fabric softeners, those used for fabric finishing, which are generally based on silicone emulsions [4-6] and household softeners, which are generally composed of cationic surfactants [7, 8]. A combination of both technologies has also been implemented recently [9]. The amphiphilic molecules used in these conditioners are double-chain surfactants that self-assemble in vesicles [10-13]. The vesicles have a dual role: they provide long-term stability and transport active ingredients such as oils [14] and fragrance capsules [15, 16] onto the fibers. Throughout the history of softeners, surfactants have been replaced by more effective and less toxic ones [8]. The development of esterquats, which are double-tailed quaternary ammonium surfactants with ester linkages making them biodegradable, is considered a critical step toward sustainability. These materials are basically produced by chemical modification of palm oil [17, 18]. Considering the growing consumer ecological awareness and the today's concern about palm oil environmental profile [19, 20], there is an ongoing research for additives with less environmental impact. In this context, apart from natural components [21, 22], the use of hydrophilic polymers can play a crucial role. Hydrophilic polymers and copolymers exist in fabric softeners. Silicone-based polymers are widely used to enhance the conditioning effect [4, 9]. Plant derived propylene glycol is considered as a biodegradable emulsifier of oils [2] included in softeners [23]. The use of cationic polymers for esterquat-free fabric softening has been mentioned recently [24].

In this direction, Solvay has recently developed a textile conditioner with reduced esterquat content, improved softening effect and fragrance delivery by including two polysaccharides in the formulation [25, 26]. These polymers originate from the seeds of cyamopsis tetragonalobus (guar gum), a legume polysaccharide, through chemical modification by cationic or hydroxyl propyl groups. The obtained modified guar biopolymers, cationic (C-Guar) and hydroxyl propyl (HP-Guar) were selected as 'green' components, since they have been characterized as safe for food and personal care products [27]. C-Guar is considered to provide smoothness and conditioning to hair [28, 29] and skin [27]. In fact, it temporarily increases the cosmetic value and function of the hair shaft until it is removed with water. The hair softening mechanism is based on the deposition of the polymer on the shafts through electrostatics, which prevents combing. Besides, C-Guar was recently shown to increase the safety of use of body wash gels [30]. HP-Guar has been extensively applied as additive for controlling the rheological properties and stability of formulations, especially foods and pharmaceutical [31] but also personal care field *i.e.* artificial tears [32, 33], cosmetic emulsions [34]. These HP-Guars properties are directly associated with the molar substitution (MS) [35].

HP-Guar and C-Guar were selected as additives for new softener formulations to allow esterquat reduction up to 50% without adverse effects on the physicochemical properties and performance of the product. The esterquat was shown to self-assemble in vesicles [36]. In order to evaluate the performance and optimize the composition of the conditioner, our group followed a new approach which allowed to study the phase behavior of the different components in the bulk. In particular, we chose to utilize cellulose nanocrystals (CNCs) as model of cotton [36-38]. CNCs are rod-like nanoparticles (~ 100 nm) derived from cellulose biopolymer which are well-dispersed in water thanks to their electrostatic charge and size.

Interaction of CNC with softener components (esterquat, HP-Guar, C-Guar) separately or in various combinations was followed by dynamic light scattering (DLS), zetametry and microscopy in the bulk and revealed strong interaction between CNC and esterquat or C-Guar due to electrostatics. Quartz crystal microbalance with dissipation (QCM-D) studies revealed that esterquat vesicles adsorb on CNC substrate and the guar biopolymers deposit with the vesicles on the CNC surfaces, the adsorption being more homogeneous and rigid in presence of polymers [39]. Our results indicated that polymers act synergistically with vesicles, enhancing the product efficiency. The performance of the guar containing formulations was tested by Solvay [25, 26]. It was found that when HP-Guar and C-Guar are added in minor quantities in the TEQ formulation, the softness efficacy is increased compared to the benchmark softener (TEQ 10 wt. %), even when TEQ is reduced at 6 wt. %. Optical fluorescence microscopy studies on cotton suggested the adsorption of intact vesicles on the fibers [40], in agreement with the literature [41]. Apart from the conditioner performance, the effect of guar biopolymers on the physicochemical characteristics of the conditioner is considered of great importance, as the insertion of additives in industrial formulations is usually challenging. Physicochemical experiments showed that guar biopolymers do not affect the properties of the vesicles but induce a depletion effect on the formulation structure in the concentrated regime [37].

Herein, we review the impact of C-Guar and HP-Guar on the esterquat vesicles and we analyze in more details their effect through dynamic light scattering (DLS), Zeta-potential, optical microscopy and cryogenic electron microscopy (cryo-TEM). Moreover, we present a methodology for visualizing the deposition of the esterquat or esterquat/guar mixtures on cotton fibers through labeling vesicles with carbocyanine dyes. Cotton fabrics treated with the softener were further investigated by scanning electron microscopy. The results presented here, together with those reported previously [36, 37, 39], show that the addition of minor amounts of natural polymers can reduce the amount of surfactants in conditioners and allow for formulations with a reduced environmental impact. This approach can facilitate the design of a class of sustainable home- and personal-care products.

## 2. Materials and Methods
**Materials**

The cationic surfactant, ethanaminium 2-hydroxyN,N-bis(2-hydroxyethyl)-N-methyl-esters, used in this work was provided by Solvay® and will be abbreviated TEQ in this manuscript. This molecule is a quaternary surfactant with C16-18 aliphatic chains and methyl sulfate anions as counterions. Its chemical structure is provided in Fig. 1a. TEQ presents a gel-to-fluid transition at 60°C due to the long-range order in the membrane [36]. Characterization of TEQ aqueous dispersions at different concentrations (0.001 – 1 wt. %) with DLS revealed the existence of particles. In particular, the second-order correlation functions $g^{(2)}(t)$ as a function of delay time in Fig. 1a display a unique relaxation mode at all concentrations. The size distribution at 0.1 wt. % presented in the same figure shows two size distributions at ~ 100 nm and ~ 1000 nm. The observed particles are attributed to vesicles as it will be analyzed in the next section. Zeta potential experiments performed on the dilute dispersions showed $\zeta$ = +65 mV, indicating highly positive particles.

The polysaccharide polymers used in this work derive from guar gum after modification with cationic or hydroxypropyl groups and were prepared by Solvay. Guar gum is a natural polymer produced through extraction from the seeds of cyamopsis tetragonalobus plant. Its monosaccharide unit is composed by β-1,4 linked mannose units and randomly attached α-1,6 linked galactose units as side chains. Guar modification by cationic groups and hydroxypropyl groups results in copolymers known as cationic (C-Guar) and hydroxypropyl guar (HP-Guar) respectively. They are consisted by monosaccharaide units of guar gum and the derivatives as shown in Fig. 1b and c. Their properties depend on the extent of substitution in the guar polymer chain. The molecular weight of C-Guar is $0.5 \times 10^6$ g mol$^{-1}$. The molecular weight of HP-Guar is $2 \times 10^6$ g mol$^{-1}$. C-Guar is characterized by a degree of substitution (DS) of ~ 0.1 and HP-Guar by a molar substitution (MS) of 0.6. In Fig. 1b and c, the second-order autocorrelation functions $g^{(2)}(t)$ of C-Guar and HP-Guar aqueous dispersions respectively, after filtration, at different concentrations (0.001 – 0.1 wt. %), are shown as a function of delay time. For C-Guar the intensity distribution exhibit one peak at ~ 200 nm (c=0.02 wt. %) while for HP-Guar a double peak with maxima at 50 and 350 nm is observed. The obtained hydrodynamic diameters are larger than the ones estimated from polysaccharides with respective molecular weights, $M_W$ = 0.5 and $2 \times 10^6$ g mol$^{-1}$ [42], indicating the possible association of the chains in water in hydrocolloid particles. Besides, guar gum is known as hydrocolloid [43]. Zeta potential studies of dilute guar dispersions showed +30 mV and 0 mV for C-Guar and HP-Guar respectively.

PKH67 fluorescent dye belongs to the family of PKH linkers which are designed for cells analysis. PKH dyes are lipophilic molecules which insert spontaneously into the cell or liposomes bilayers. Their photophysical properties cover the visible and infrared emission spectrum [44]. PKH67 consists of two fluorescent head groups and two aliphatic chains. It fluoresces in the green, the excitation wavelength is at 410 nm while the emission one at 502 nm. PKH67 was purchased from Sigma-Aldrich in 1 mM ethanol dispersion.

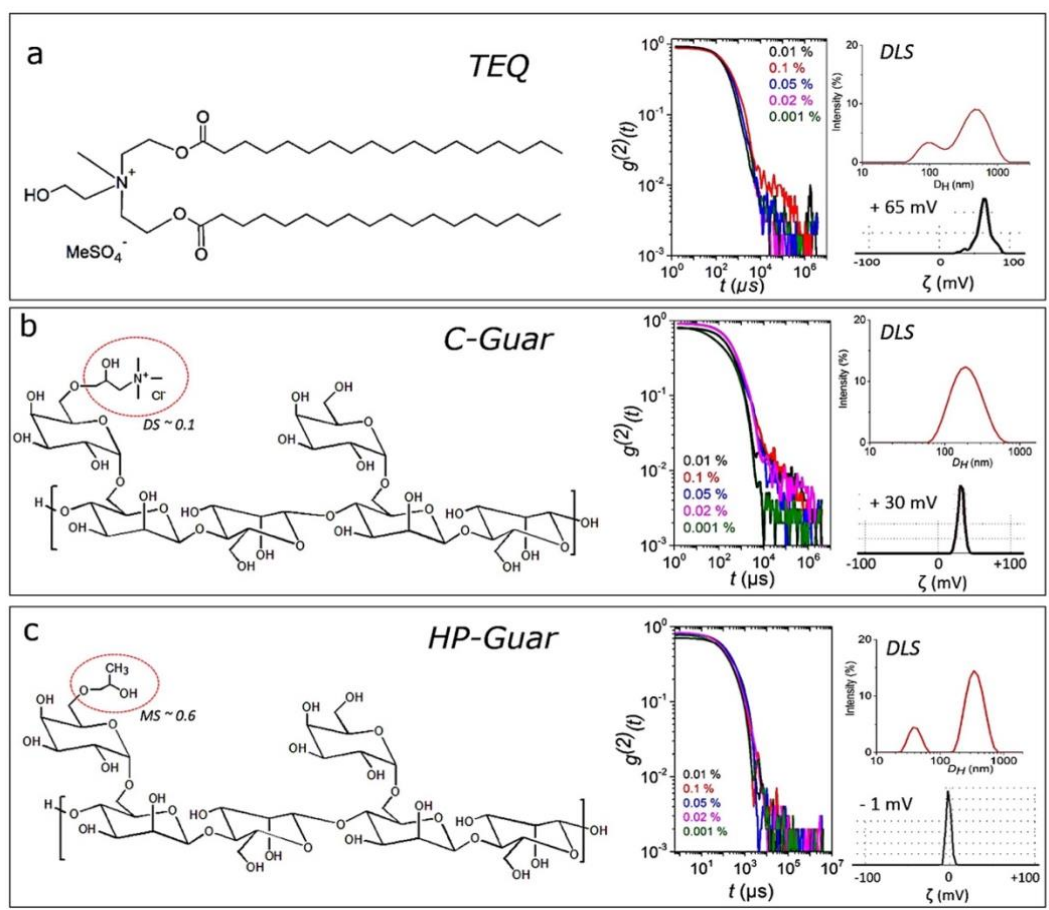

*Figure 1. Chemical structure, second-order auto-correlation functions, size distribution and ζ-potential of (a) TEQ surfactant, (b) C-Guar and (c) HP-Guar aqueous dispersions.*

**Sample preparation**

A dispersion of TEQ at 6 wt. % was prepared by adding dropwise melted TEQ in filtered MilliQ water at 60 °C. The pH was fixed at 4.5. Formulations containing TEQ 6 wt. % and the biopolymers were prepared by dispersing the guars in 60 °C water, adjusting the pH at 4.5 and adding the melted TEQ at the final step. The guar concentration was chosen to be 0.4 wt. % when only one guar was used and 0.3 wt. % for each biopolymer for the formulations containing both C- and HP-Guar. Formulations with TEQ 4 wt. % were prepared for SAXS experiments. When diluted solution were studied, the dispersions were prepared by mixing the concentrated ones with MilliQ filtered water (T = 25 °C). 1 wt. % stock solution of each guar were prepared at 60 °C adjusting the pH at 4.5.

**Deposition of labeled TEQ vesicles on cotton fabrics**

10 μl PKH67 were mixed with 1500 μl DI-water rapidly using vortex. The obtained dispersion was then mixed with 1500 μl of 1 g L$^{-1}$ TEQ or TEQ in presence of guar biopolymers using vortex. Cotton fabrics cut in small pieces of 0.03 g were immersed for 10 minutes in the dispersion. For studies of dried cotton, the treated cotton pieces were placed at 30 °C under air circulation for 2 hours.

**Methods**

**Light scattering**. A Zetasizer Nano ZS spectrometer (Malvern Instruments, Worcestershore, UK) was used for light scattering experiments. For the illumination, a 4 mW He–Ne laser beam ($\lambda$ = 633 nm) is applied while the scattering angle is fixed at 173°. Stokes-Einstein relation, $D_H = k_B T/3\pi\eta D_C$, where $\eta$ the solvent viscosity, $k_B$ is the Boltzmann constant and $T$ the temperature was applied to determine the hydrodynamic diameter. Cumulant along with CONTIN algorithms were used for determining the average diffusion coefficient $D_C$ of the scatterers at the second-order autocorrelation function, $g^{(2)}(t)$. The experiments temperature was fixed 25 °C and all measurements were conducted after 120s (equilibration). All experiments run in triplicate.

**Zeta Potential.** The zeta potential and electrophoretic mobility of the diluted samples were studied with a laser Doppler velocimetry (Zetasizer, Malvern Instruments, Worcestershore, UK) applying an angle of 16° and using the phase analysis light scattering mode. The experiment temperature was set at 25 °C, and the samples were let 120 s for thermal equilibration. All experiments run in triplicate.

**Single angle X-ray scattering**. 1.5 mm quartz capillaries purchased from Glass Müller were used for Small-Angle X-ray Scattering (SAXS) experiments. Paraffin was added on the top of them to avoid water evaporation. SAXS studies were conducted on the SWING beamline at the SOLEIL synchrotron source (Saint-Aubin, France) using a two-dimensional AVIEX CCD detector which was placed in a vacuum detection tunnel. 0.08 – 8 nm-1 $q$-range was the distance between the sample and the detector and 12 keV was the beamline energy used. The temperature of the experiments was set at 25 °C. Five acquisitions were performed for each sample and averaged; 250 ms was the acquisition time for each pattern. Silver behanate was used for the calibration. The scattered intensity was recorded as a function of the scattering vector $q = 4\pi sin\theta/\lambda$, where and $\lambda$ is the wavelength of the incident beam and $2\theta$ the scattering angle. A normalization of the intensity values was performed in order to account for intensity beam, sample transmission and acquisition time. Every scattering pattern was integrated circularly to define the intensity related to the wave-vector. A sample with water was studied and its scattered intensity was then subtracted from each sample scattering record.

**Optical microscopy.** An IX73 inverted microscope (Olympus) was used for phase contrast optical microscopy studies. 20×, 40×, and 60× objectives were used for the experiments. The images and videos were obtained by using an Exi Blue camera (Qimaging) along with a Metaview software (Universal Imaging Inc.). ImageJ software and plugins (http://rsbweb.nih.gov/ij/) was used to treat the obtained images.

Seven microliters of TEQ dispersions and TEQ/guar mixtures were sealed into a Gene Frame (Abgene/ Advanced Biotech) dual adhesive system after deposition on a glass plate. For studding the cotton treated with labeled TEQ, a piece of cotton was placed into the same Gene Frame. For observations performed in water, 30 µl water were added around the cotton piece.

**Cryogenic transmission electron microscopy (Cryo-TEM)**. A TEM microscope (JEOL 1400 operating at 120 kV) was used for cryo-TEM studies. Liquid nitrogen temperature was obtained by using a cryo holder (Gatan). A 2k-2k Ultrascan camera (Gatan) was used. Magnifications between 3000× and 40000× were applied. For the sample preparation, 7 microliters of the sample were deposited on a lacey carbon coated 200 mesh (Ted Pella). VitrobotTM (FEI) was used for blotting the sample by a filter paper. The grid was then immersed in liquid ethane and finally transferred to the vacuum column of the microscope.

**Rheology**. A Physica RheoCompass MCR 302 (Anton Paar) equipped with a cone-and-plate geometry (cone angle 1°, diameter 50 mm), was used for studding the rheological properties of the formulations. Water evaporation was avoided due to the use of a solvent trap. The experiment temperature was fixed at 25 °C. The storage and loss moduli, $G'(\omega)$ and $G''(\omega)$, were studied as a function of the frequency by applying an oscillatory shear stress of 1 rad s$^{-1}$. Before, the linear viscoelastic regime was defined by strain sweep experiments at a frequency of 1 rad s$^{-1}$.

**Scanning Electron Microscopy (SEM).** Woven cotton fabrics 5 x 5 cm were used for studying the deposition of the softener onto the fibers. The fabrics are made of cotton yarns (200 – 350 µm) consisted of cotton fibers (D = 10 – 20 nm). The fabrics before the treatment with the softener formulations, were immersed in water (Milli-Q) for 10 min and were dried at 35°C. Then, they were immersed in 100 mL of the softener solution for 10 min under stirring. After this treatment, the fabrics were deposited horizontally without any contact and were dried at 35 °C under air circulation for one hour. Two solutions were used for the treatment: TEQ 1 wt. % and TEQ 1 wt. %/C-Guar 0.03 wt. % /HP-Guar 0.03 wt. %. For SEM observation SUPRA 40V SEM-FEG (Zeiss, Oberkochen, Germany) was used. In-lens secondary electrons detector and 2 kV acceleration tension were used in 2-3 mm working distance. All materials were metallized: a 2 nm layer of platinum was deposited on fabrics.

## 3. Results and Discussion
### 3.1 Characterization of vesicles aqueous dispersions

Before analyzing the self-assembly of TEQ surfactants in vesicles, we describe here the most common types of vesicles based on their lamellarity, a term that defines the number of closed bilayers that compose them. This classification is presented in Fig. 2a. Multilamellar vesicles or onions are multiple concentric lipid bilayers encapsulating an aqueous compartment; multivesicular vesicles encase smaller vesicles on different sizes, while unilamellar vesicles are composed of a single bilayer [46]. A critical tool to visualize surfactant or lipid vesicles in water is cryogenic transmission electron microscopy which allows for high-resolution structure determination of biomolecules in solution. A limitation of this technique for vesicular dispersions is the vesicle size and concentration [47, 48]. Indeed, the vitrified water layer between the cryo-TEM grids has a thickness which varies generally between 100 and 400 nm. Hence, larger vesicles are excluded during the preparation and they are not detected by this technique, unless they are deformed and flattened, making them appear non-spherical. Such vesicles are schematically represented in Fig. 2b.

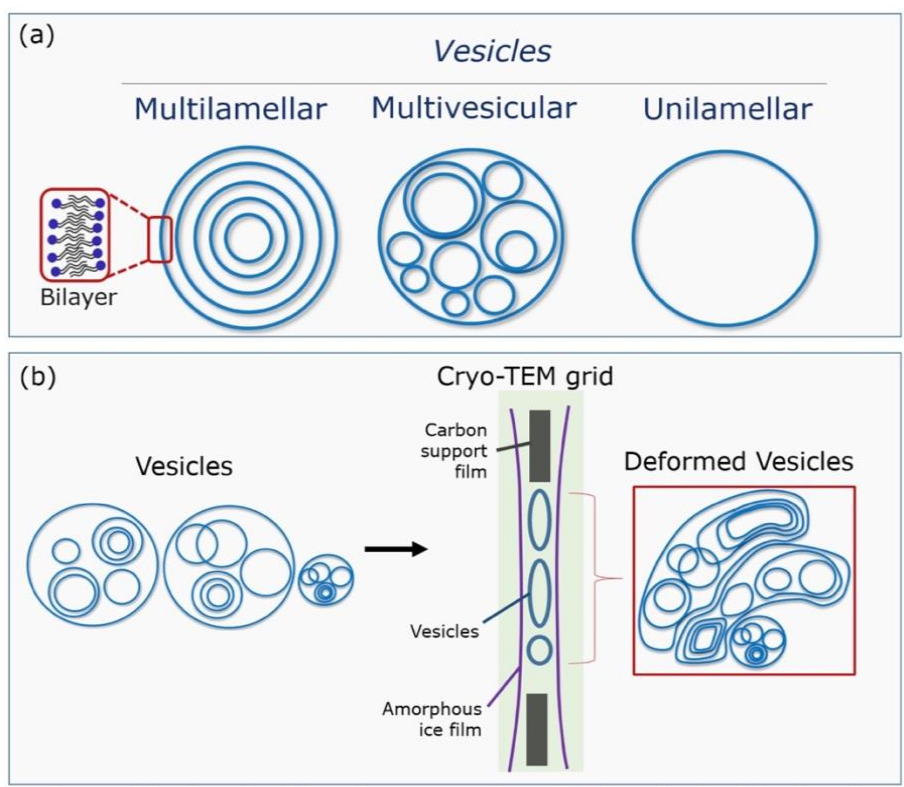

*Figure 2. (a) Classification of vesicles according to their lamellarity, (b) Schematic representation of vesicle deformation obtained by cryo-TEM when their sizes are larger than the vitrified water layer thickness.*

Deformed vesicles are identified by cryo-TEM for TEQ aqueous dispersion at concentrations c = 4 wt. % and 1 wt. % in Fig. 3a and b respectively. At 4 wt. %, the majority of vesicles are larger than 200 nm and are therefore flattened from the confined geometry of the amorphous ice film. At both 4 wt. % and 1 wt. %, multivesicular vesicles are the most common structures. Other structures such as doublets and stomatocytes (concave shapes) are also present in minute proportions at 4 wt. % and are indicated in Figs. 3a,b. Seth *et al.* have ascribed those structures to the electrostatic induced destabilization occurring for crowded and charged vesicles [49]. Multilamellar vesicles are scarcely found. In the diluted regime (c = 0.1 wt. %, Fig. 3c) unilamellar vesicles of size 50 to 200 nm prevail. Cryo-TEM data thereby reveal that the vesicles are stable upon dilution. Note here that the conditions of use in the washing machine correspond to ~ 0.01 wt. % while stocking conditions for the guar containing formulation is 6 wt. %. The thickness of the bilayer has been assessed by high magnification image analysis to be 5.2 ± 0.6 nm [36], in agreement with two-dimensional assembly of 16-18 carbon atom surfactants.

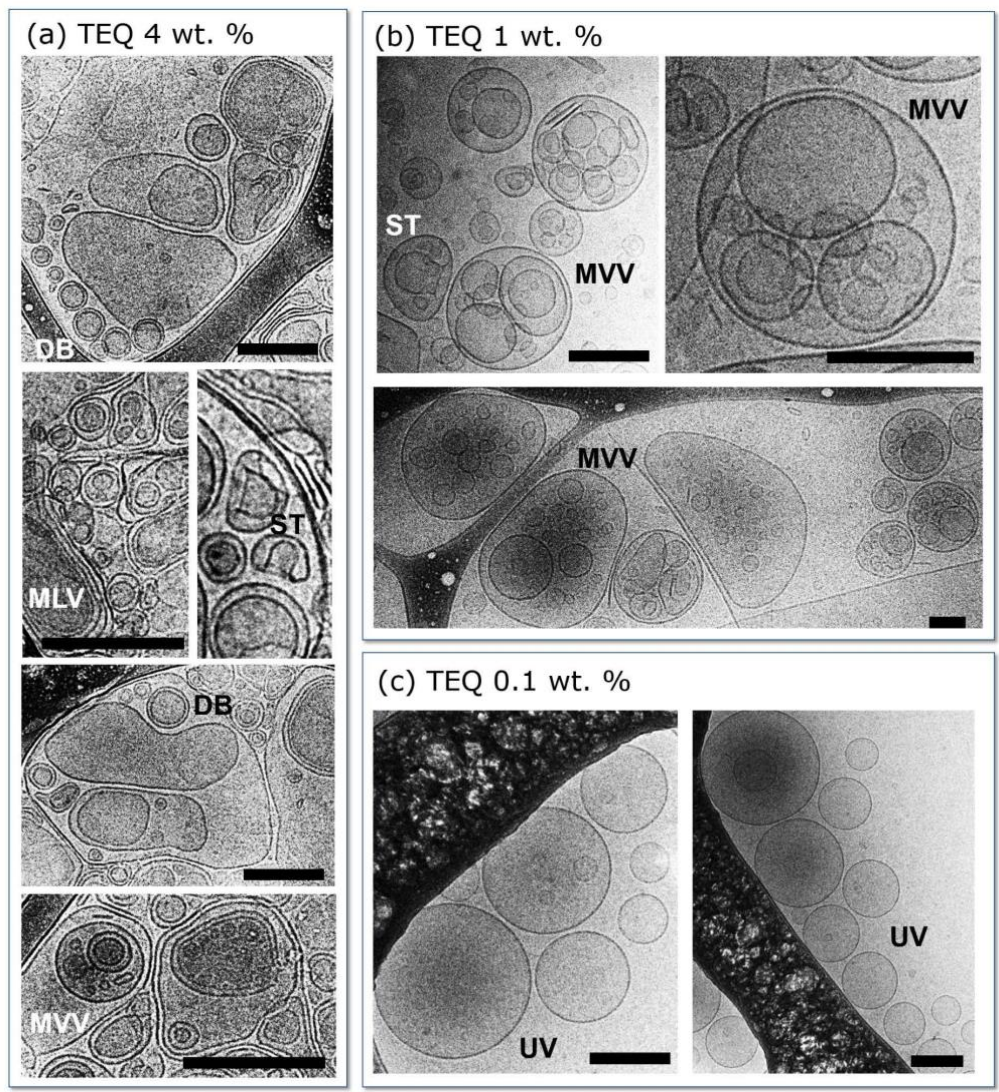

*Figure 3.* Representative cryogenic transmission electron microscopy (cryo-TEM) images of (a) 4, (b) 1 and (c) 0.1 wt. % TEQ aqueous dispersions. Multivesicular (MVV), doublets (DB), unilamelar (UV), stomatocytes (ST), multilamellar (MV) and unilamellar (UV) vesicles are observed. Scale bars are 200 nm.

## 3.2 Characterization of vesicles dispersions in presence of guar biopolymers

### 3.2.1. Microscopy

Representative cryo-TEM images of 4 wt. % TEQ in the presence of C-Guar are shown in Fig. 4a. As with 4 wt. % TEQ, multivesicular vesicles are predominant after the addition of C-Guar. However, in contrast to TEQ 4 wt. %, the vesicles are smaller and appear to be aggregated, probably resulting from polymer-mediated attractive depletion interaction [50]. In the presence of both C- and HP-guars however (Fig. 4b), the vesicles are larger and less aggregated. This indicates that C-Guar causes vesicle crowding effects, while the addition of HP-Guar makes the dispersion more homogeneous. These data show that although the vesicles are stable upon guar addition, the overall structure being preserved, there are significant differences in vesicle organization and size compared to the TEQ dispersions shown in Fig. 4a.

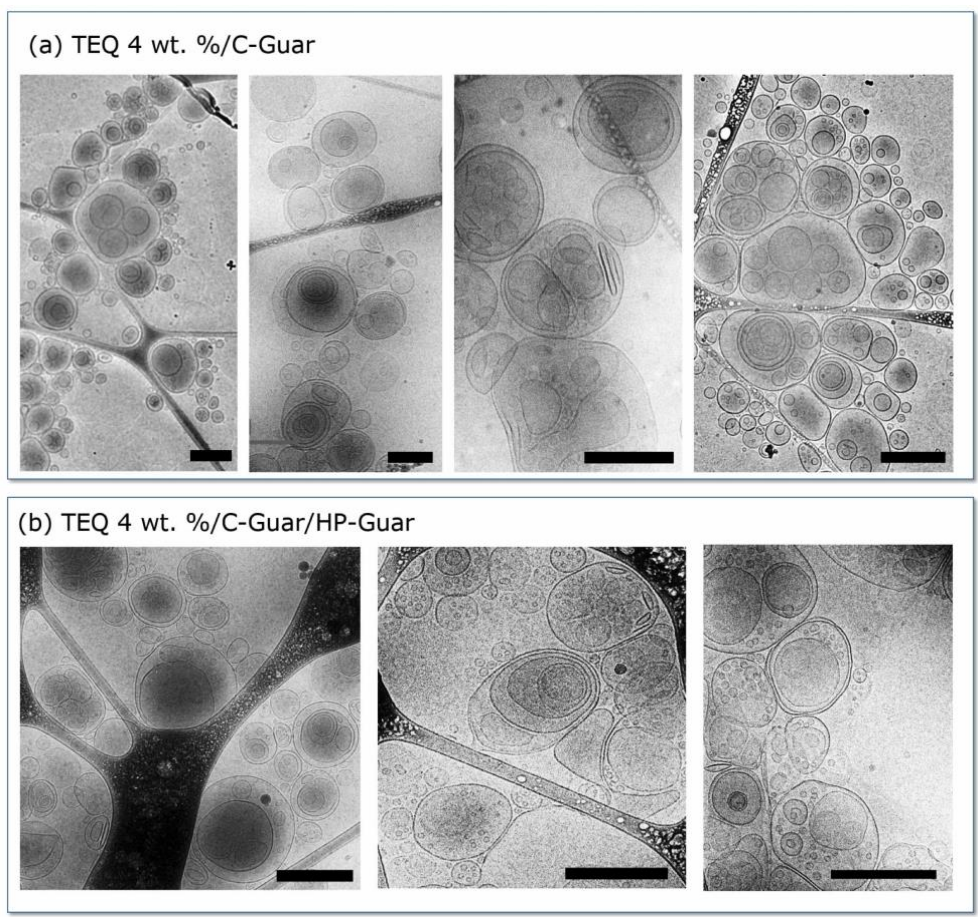

*Figure 4.* Representative cryogenic transmission electron microscopy (cryo-TEM) images of TEQ 4 wt. % in presence of (a) C-Guar and (b) mixture of C-Guar and HP-Guar. Scale bars are 500 nm.

Differences induced by the guar addition on TEQ vesicle dispersion were further examined by phase contrast optical microscopy [51]. In Fig. 5a, a 20x magnification image of a 4 wt. % TEQ aqueous formulation is exhibited. We observe, spherical (green arrows) or faceted (red arrows) well contrasted objects, defined as micron-sized vesicles. The faceted vesicles are likely due to mechanical stresses generated during the formulation and which remain after their formation. The faceted vesicles have been previously assigned to the organization of the alkyl chains into the bilayer in a hexagonal liquid-crystalline order, as revealed by wide-angle X-ray scattering experiments [52, 53].

The formulation is characterized by a size distribution centered on 1 µm and by a large dispersity. Some vesicles are larger than 5 µm (yellow arrows). Time-lapse video-microscopy showed that the particles exhibit rapid Brownian motion, attesting to the existence of a free volume for the vesicles even at these concentrations [36]. The addition of minor quantity of C-Guar (0.4 wt. %) induced the aggregation of the vesicles (Fig. 5b). Compared to Fig. 5a, no vesicles larger than 2 µm are observed under these conditions. These results are in agreement with cryo-TEM observations (Fig. 4b) where we found that the vesicles are aggregated and smaller than in absence of guar biopolymers. In addition, the vesicles in presence of C-Guar do not undergo Brownian motion as in TEQ 6 wt. %. Adding a mixture of C-Guar and HP-Guar to a 6 wt. % TEQ formulation results in a different structure and morphology, as illustrated in Figure 5c. We find here a heterogeneous micron-sized texture, frozen at the time scale of the observation, (of the order of the minute). The impact of each guar on the morphology of the formulation is critical at the stock condition concentration.

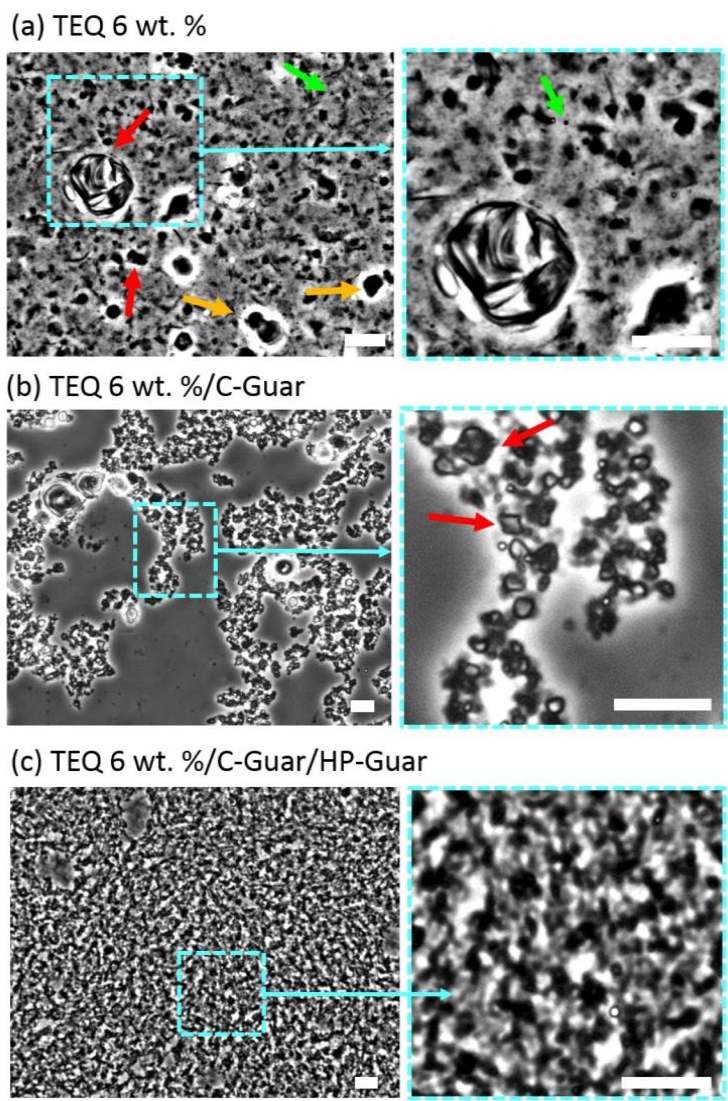

***Figure 5.*** *Phase contrast microscopy images of 6 wt. % TEQ dispersions (a) without guar biopolymers, (b) in presence of 0.4 wt. % C-Guar and (c) mixture of C-Guar and HP-Guar (0.3 wt. % each). Red, green, yellow arrows indicate faceted, spherical and large vesicles respectively. Scale bars are 20 µm.*

Similar studies have been undertaken at low surfactant concentrations, typically 0.1 wt. % which is close to the concentration used in washing machines, and the effects of C-Guar and HP-Guar could not be clearly identified. With cryo-TEM, formulations in the absence (Fig. 6a) or presence of C-Guar and HP-Guar (Fig. 6b) show mainly unilamellar vesicles in the 50-200 nm range. Only a few multivesicular and multilamellar vesicles are observed in such conditions. Phase contrast optical microscopy studies show well-contrasted and non-aggregated particles which are animated by Brownian motion. The size distribution of these particles is peaked around 0.8 µm, in accordance with what was found in paragraph 3.2.1. These findings can be explained by the existence of depletion interactions between vesicles induced by polymers, such interactions having a significant effect only at higher concentration.

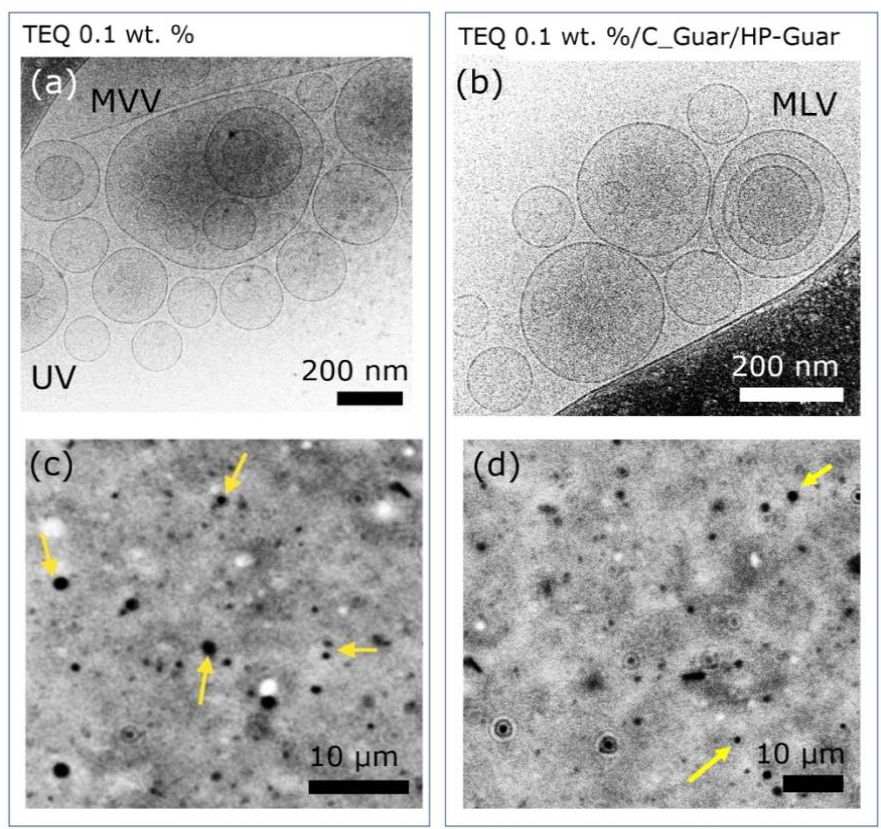

*Figure 6.* TEQ 0.1 wt. % representative cryogenic transmission electron microscopy (cryo-TEM) (a), (b) and phase contrast microscopy images (c), (d), in the absence (a) (c) and in the presence of guar polymers (b), (d). Unilamellar (UV), multivesicular (MV) and multilamellar vesicles (MLV) are observed.

3.2.3 X-Ray scattering

The impact of guar biopolymers on vesicle properties was further examined by means f small angle X-ray scattering. The scattering cross-section of the two guar biopolymers at c=0.6 wt. % are shown in Fig. 7a. A decrease which is expected for polymer solutions, especially a scaling law recorded at large wave-vectors (q > 0.3 nm$^{-1}$) is observed. However, the q-range studied does not allow the determination of the radius of gyration of the polymers, an outcome that can be imputed to the dispersity of the chains. In contrast, for TEQ 4 wt. %, the intensity shows a broad oscillation which can be interpreted in terms of the form factor $P_{BL}(q)$ of a surfactant bilayer [52, 54], in the range 0.4 - 2.5 nm$^{-1}$. The oscillation is here attributed to the variation of electron scattering density across the bilayer thickness. Depicted in Fig. 7b, the two-level electronic density profile depends on two parameters, the size of the head groups $\delta_H$ and the thickness of the hydrophobic region $2\delta_{ST}$. Fig. 7c displays the SAXS intensity for TEQ 4 wt. % in linear scales, together with the adjustment of the bilayer form factor [52, 54] using as fitting parameter $\delta_H$ = 0.8 ± 0.05 nm and $2\delta_{ST}$ = 3.0 ± 0.1 nm. Here, we used the expression provided by Kawabata *et al.* [52]:

$$P_{BL}(q) = \frac{4}{q^2}\left[(\rho_H - \rho_W)\sin(q(\delta_H + \delta_{ST})) + (\rho_{ST} - \rho_H)\sin(q\delta_{ST})\right]^2 \quad (1)$$

where $\rho_W$, $\rho_H$ and $\rho_{ST}$ are the scattering length densities of water, hydrophilic and hydrophobic parts of surfactants, respectively. The total bilayer thickness can hence be estimated at 4.6 ± 0.15 nm. This value is consistent with the results obtained before [36] by cryo-TEM showing a thickness of 5.2 ± 0.1 nm. After the addition C-Guar, HP-Guar and C-Guar/HP-Guar, the TEQ form factor is not changed, indicating that there is no significant modification in the TEQ bilayer. Interestingly, low amplitude Bragg peaks superimposed to the bilayer form factor and indicated by arrows are observed in the three conditions [52]. For TEQ/C-Guar, the Bragg peaks are located at 0.6, 0.9 and 1.18 nm$^{-1}$ whereas for TEQ/HP-Guar they are found at 0.62, 0.91 and 1.14 nm$^{-1}$. With the combination of the two guars (at 0.3 wt. % each), the Bragg peaks are still present and located at 0.74, 0.98 and 1.15 nm$^{-1}$, i.e. close to the previous values. The peaks obtained for TEQ/guar mixtures are similar to those obtained for the concentrated TEQ dispersions at 8, 12 and 15 wt. % [36]. For these dispersions, the superimposed peaks were assigned to structure factors of closely packed surfactant membranes which are encapsulated into multilamellar vesicles. Thus, adding guar biopolymers or increasing the TEQ concentration has a similar effect on the multivesicular surfactant structure. These results were attributed to crowding effects associated to depletion triggered from the polymers along with the overall increase of the vesicular volume fraction.

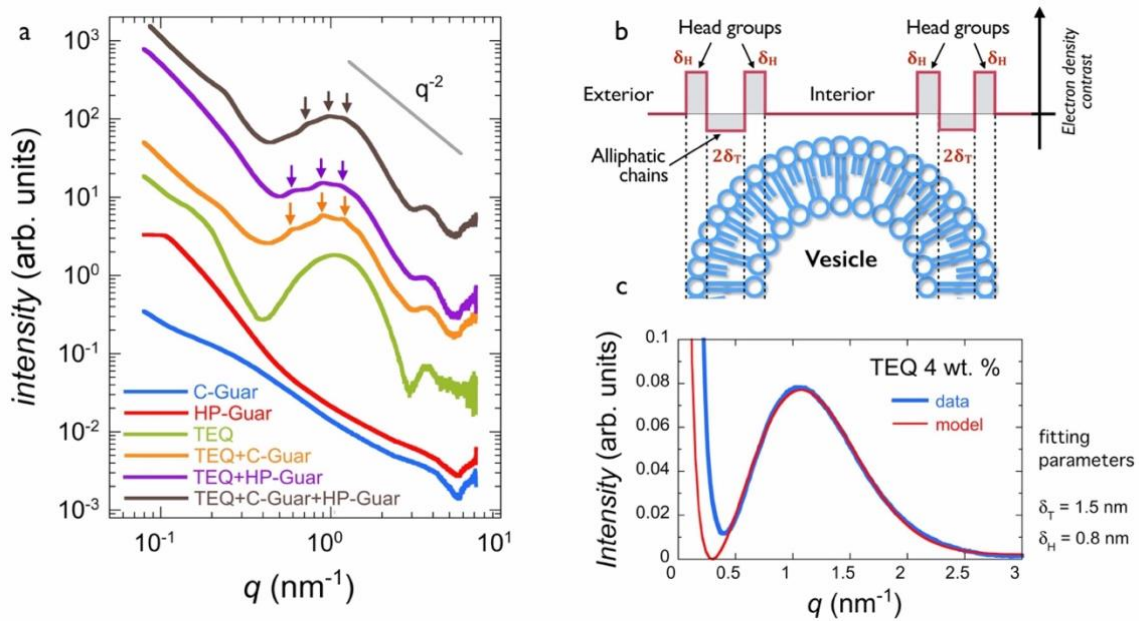

*Figure 7.* (a) Small-angle X-ray scattering from C-Guar, HP-Guar at 0.6 wt. %, TEQ mixed with C-Guar (0.4 wt. %), HP-Guar (0.4 wt. %) or mixture of C- and HP-Guar ($c_{tot}$= 0.6 wt. %). TEQ concentration is fixed at 4 wt. % for all experiments. The arrows show the Bragg peaks. (b) Schematic representation of the two-level electronic density profile across the vesicle bilayer.

*There $\delta_H$ denotes the size of the head groups and $2\delta_{ST}$ the thickness of the hydrophobic region. (c) Small-angle X-ray scattering from TEQ along with adjustment of the bilayer form factor*

3.2.1. Rheology

Rheological characterization of C-Guar, HP-Guar and C-Guar/HP-Guar aqueous dispersions at concentration 0.4 wt. % shows viscous polymeric fluids (Figs. 8a-c). HP-Guar and the mixture of both guar biopolymers exhibit slight viscoelastic behavior ($G''(\omega) > G'(\omega)$). Such rheological characteristics arise from entanglements attributed to intermolecular association in combination with H-bonding between the polymer chains (hydrocolloids) [35]. The similarity of the curves in Figs. 8b and c indicates that the rheological properties of the mixture are mainly defined by the HP-Guar.

The addition of guar to a TEQ dispersion significantly alters its behavior. In particular, a 6 wt. % TEQ formulation shows liquid-like behavior (*i.e.* $G''(\omega) > G'(\omega)$), with $G''(\omega)$ and $G'(\omega)$) ~ 0.01 Pa (Fig. 8d). C-Guar addition triggers both moduli increase while $G'$ is now higher than $G''$ suggesting a gel-like behavior typical for soft solid materials. When HP-Guar or mixture of guar biopolymers is used, the moduli are further enhanced. Also, guar biopolymers make the TEQ formulation a non-zero yield stress material, the yield stress being around 0.1 Pa as shown in Fig.8e. In Figs 8d and 8e it is found that the rheological properties are mainly determined by the HP-Guar, in agreement with the observations made for solutions without TEQ (Figs. 8 a-c). The results obtained for TEQ/guar mixtures are similar to the rheological characteristics of TEQ 10.5 wt. % [37]. Hence, as suggested by x-ray studies, TEQ concentration or addition of guar biopolymers have comparable effects to the vesicle dispersions. The viscosifying properties induced by guar biopolymers support our suggestion for depletion phenomena due to crowding and vesicle aggregation. Besides, additional role is played by the thickening characteristics of guar biopolymers. Homogeneity of the formulation despite these phenomena in presence of HP-Guars is achieved possibly thanks to steric effects of the hydroxypropyl groups along the polymeric chain. In conclusion, the role of the guar biopolymers is crucial for the rheological behavior of the formulation, which is critical for the product stability and also for consumer appeal.

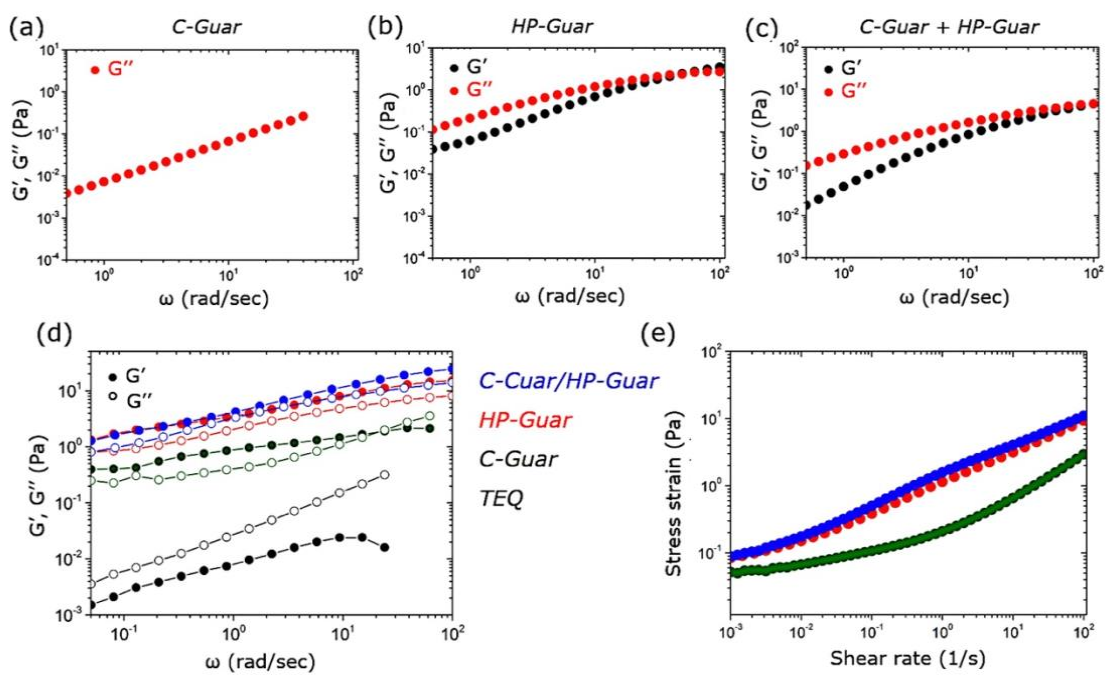

**Figure 8**. Elastic and loss modulus $G'(\omega)$ and $G''(\omega)$ as a function of the angular frequency obtained from (a) C-Guar 0.4 w. %, (b) HP-Guar 0.4 wt. %, (c) mixture of guars ($c_{tot}$: 0.4 wt.%). Elastic and loss modulus $G'(\omega)$ and $G''(\omega)$ (d) and shear stress (e) obtained from TEQ 6 wt. %, TEQ 6 wt. % mixed with C-Guar (0.4 wt. %) or HP-Guar (0.4 wt. %) and guar mixture ($c_{tot}$: 0.6 wt. %). The results have been obtained by rotational cone-and-plate rheometry.

### 3.3. Adsorption studies on cotton fabrics

Fig. 9a shows a SEM image of the cotton fabric used for studding softener deposition on cellulose fibers. It is a woven fabric made by two sets of 250 µm spun yarns interlaced at right angles with respect to each other [55]. The yarns consist in interlocked 10 – 20 µm cotton fibers and are separated by large voids (arrows). To elucidate the deposition mechanism of vesicles onto cellulose fibers, we proceeded to fluorescent microscopy. Before, we stained the vesicles to make them fluorescent (Fig. 9b). Although various techniques for labeling cells, exosomes or synthetic liposomes exist in the literature, labeling of preformed vesicles, like the vesicles used in softeners is not straightforward. In particular, such formulations need to be labeled directly without altering their concentration and structure.

Hence, we developed a straightforward methodology for staining preformed vesicles with a fluorescent molecule which is commonly used in biology, the PKH67. Its chemical structure is given in Fig. 9c. This dye is known to incorporate in the lipid bilayer of cells or liposomes. The proposed procedure was shown to appropriately stain aqueous dispersion of TEQ, directly in the formulation, without purification steps and without modifying the vesicles properties [40]. In Fig. 9b, TEQ vesicles treated with PKH67 appear as well-contrasted objects or as bright spots by phase contrast and fluorescent microscopy respectively. By comparing these images one can observe that the spots are co-localized. For instance, the yellow arrows in the two images point out the same vesicles captured either by phase contrast or by fluorescent microscopy. The great majority of the vesicles appear also as fluorescent, meaning that the vesicles have been successfully labeled without changing their overall morphology. In a previous study [37], we checked the adsorption of labeled vesicles on cellulose nanocrystals (CNC) which were chosen as model of cotton (cellulose) [36, 38]. The CNC fibers were shown to adsorb onto the TEQ vesicles in water. This procedure has been further confirmed by quartz crystal microblance (QCM) [39]. Such deposition of intact vesicles was viewed also on cotton fibers through treatment of cotton fabrics with aqueous dispersion of labeled TEQ.

Observations of these cottons submerged in water are given in Fig. 9d. By phase contrast microscopy only the woven structure is visible. In the emission mode, the same fibers appear brighter while bright spots assigned as fluorescent vesicles are indicated by the red arrows around the fibers. These observations suggest that both supported vesicular layer (SVL) (bright spots) and supported lipid bilayer (SLB) (fluorescent fibers) are obtained. Taking into account previous studies on the interaction of TEQ with CNC [36], we attribute the intense deposition of the vesicles on fibers to electrostatic interaction. Control experiments conducted with unlabeled TEQ and have been shown previously [40] indicated that the cotton fibers appear dark and no fluorescent signal is captured. Interestingly, after drying, no individual objects are detected on the fibers: only fluorescent fibers are observed, indicating that in absence of water the vesicles rupture to form one or more bilayers wrapping the fibers. To our knowledge, this deposition mechanism (in wet and dry state) has never been observed before. We suggest therefore that the deposited bilayers endow the fibers with softness and antistatic properties. Similar behavior is observed in presence of guar polymers (Fig. 9e). Bright particles attributed to vesicles are visible when the fibers are examined in wet conditions. Here, along with the vesicles, some patches (yellow arrows) assigned to guar polymers are identified. After drying, only fluorescent fibers and patches stuck on the fibers can be seen. These findings are consistent with results presented before which show the strong interaction of guar polymers with CNC [37, 39]. Our proposed mechanism showing the deposition of softener vesicles along with bilayers onto cotton fibers in water (conditions in washing machine) and the vesicles transformation to bilayers after drying, is schematically represented in Fig. 9f. In conclusion, the proposed methodology based on fluorescent microscopy allow for the visualization of the product deposition and the evaluation of the softener performance and could be applied to elucidate the adsorption of other household or personal care products on solid surfaces.

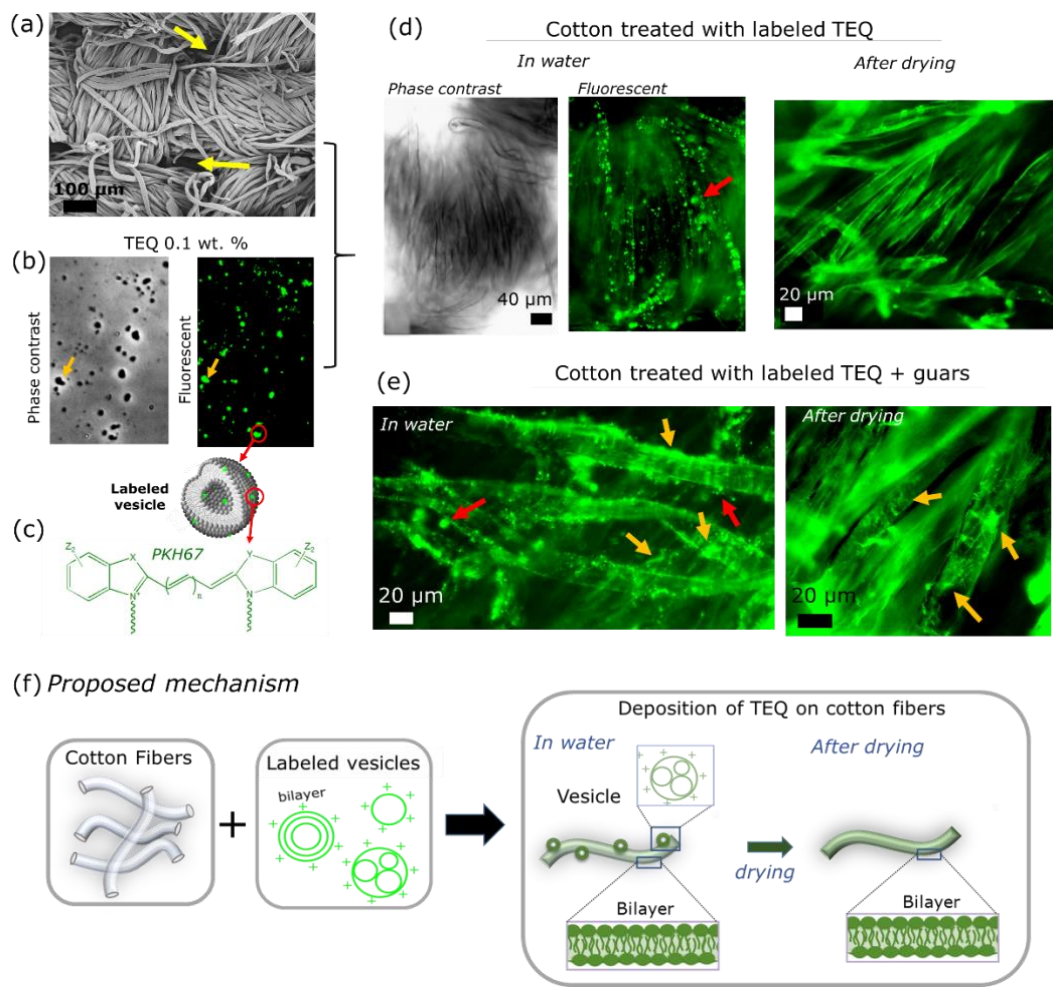

*Figure 9*: (a) 20× phase-contrast microscopy image of a woven cotton fabric, (b) phase contrast and fluorescent microscopy image of labeled TEQ 0.1 wt. % aqueous dispersion, (c) chemical structure of PKH67 amphiphilic dye consisted by fluorescent head groups $Z_1$ and $Z_2$ which emit at 502 nm and two aliphatic chains that incorporate into the membrane, (d) 40× phase-contrast and fluorescent microscopy image of a woven cotton fabric treated with labeled TEQ vesicles examined in water and in the dry state, (e) 40× fluorescent microscopy image of a woven cotton fabric treated with labeled TEQ vesicles and guar biopolymers examined in water and in the dry state, (f) schematics of TEQ formulation adsorbed on cotton fibers: both supported vesicles and supported lipid bilayers are assumed to be deposited on cotton in water and supported bilayers in the dried state.

In Fig. 10 we follow the deposition of the softener formulation on cotton fabrics by SEM. Centimetre large pieces of fabric were immersed in the softener solution ($c_{TEQ}$ = 1 wt. %) for 10 min under stirring and then dried at 35 °C for one hour. This high concentration was chosen to enable the surfactant visualization. Fig. 10a shows a SEM image of untreated cotton fibers. The fibers appear as elongated cylinders of diameter 10-20 μm and exhibit crenulations and convolutions at their surface [56]. The primarily wall of these fibers consists in numerous fine threads of cellulose called fibrils. The fibrils spiral around the fiber axis with and angle of about 70 degree, as indicated by the yellow arrows in Fig. 10a. For fibers treated with 1 wt. %

TEQ (Fig. 10b and 10c), the fibrils are not visible, indicating that a surfactant layer has been deposited on them. The smoothness of the air-cotton interfaces (blue arrows in Fig. 10c) suggests in addition that the surfactant is adsorbed not in the form of vesicles, but likely in the form of surfactant bilayers. From the data it is difficult however to infer the thickness of the coating and whether this layer is uni- or multilamellar. With guar, deposited materials are directly visible onto and between the fibers and appear as filaments connecting neighboring fibers (red arrows in Fig. 10 d,e). In some cases, the material linking the fibers appear as thin suspended films (green arrow). It is assumed that the deposited material seen at the fiber surfaces is a complex mixture of polymers and. These overall findings agree with the experiments discussed previously [37] with the cellulose nanocrystals and attest the strong affinity of guar-surfactant formulation for cotton. Also, SEM results corroborate the conclusions taken out from fluorescent microscopy where 'patches' or films attributed to guar biopolymers were found to be deposited on the fibers.

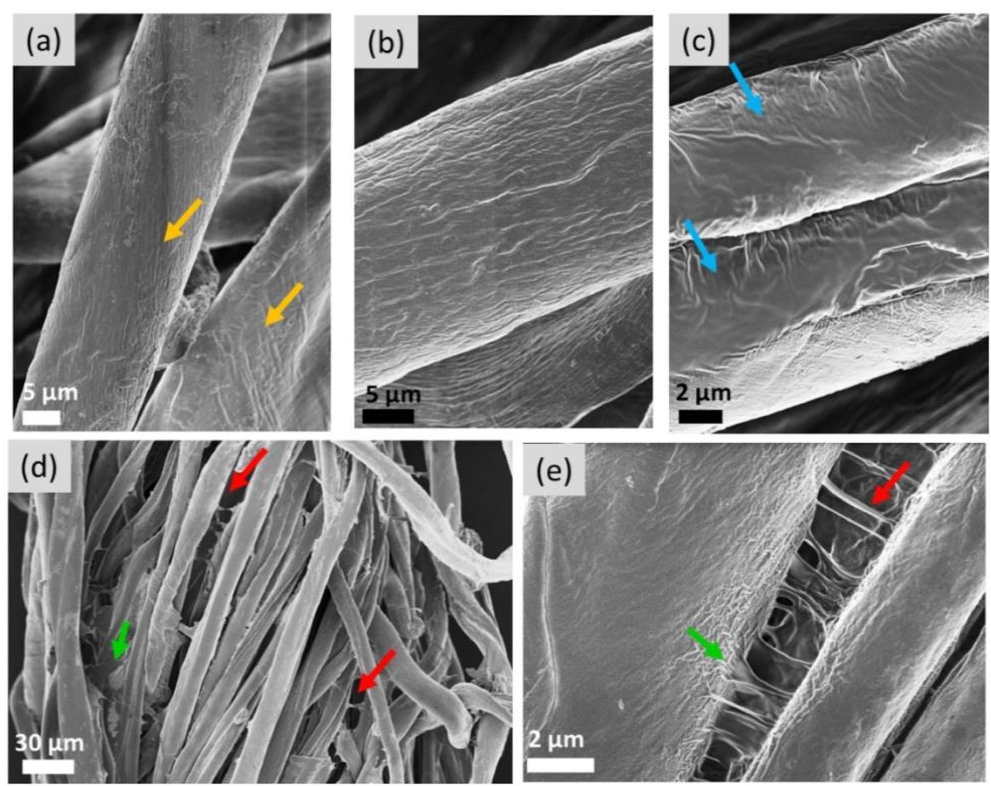

*Figure 10.* Representative SEM images of cotton fibers treated with (a) water, (b), (c) TEQ 1 wt. %, (d), (e) TEQ 1 wt. % in presence of guar biopolymers. The magnifications used in a-d) are 90, 2000, 2000, 400 and 5000 respectively. Yellow, blue, green and red arrows indicate the fibrils, TEQ layer, polymer threads and TEQ/polymer films respectively.

## 4. Conclusion

To design household products with a lower impact on the environment, we study here the possibility of reducing the concentration of surfactants in fabric softeners. In the proposed formulation more than 50 % of the surfactant, which is an C16-18 aliphatic chain esterquat derived from palm oil, has been replaced by minor amounts of two guar biopolymers, a cationic guar (C-Guar) and a hydoxypropyl guar (HP-Guar). These biopolymers are widely used as additives in the personal care and food industries. The surfactant has been shown to self-assemble in vesicles in water. Optical microscopy, cryo-TEM, SAXS and rheology studies reveal that the biopolymers have synergistic effects on the softener formulation. The vesicles are stable and preserve their overall structure when both guars are used. In the concentrated regime, SAXS indicates a transformation from multi-vesicular to multilamellar vesicles in presence of the biopolymers, attributed to crowding effects probably induced by depletion attractions. C-Guar is considered to enhance deposition on fibers due to its cationic charge. The role of HP-Guar is crucial for maintaining stability of the formulation despite these interactions. In addition, HP-Guar, thanks to its viscosifying characteristics, improve the rheological properties of the formulation by compensating for the decrease in viscosity caused by the surfactant reduction.

Deposition studies performed by fluorescent microscopy and SEM show a high affinity of the formulation with cotton fibers, again attributed to electrostatic attraction. A new protocol developed for staining preformed vesicles and observing their adsorption on fibers is presented here. This technique allows for the visualization of the fluorescent vesicles on cotton, in presence or not of biopolymers, confirming the deposition of intact vesicles on cellulose in water. Upon drying, the vesicles transform into bilayers covering the fibers, while large patches of deposited materials are found in presence of guar biopolymers. SEM confirmed the adsorption of polymers on the fibers and the absence of vesicles. The use of biopolymers presented here for replacing the main components of a formulation, along with the proposed methodologies for studding deposition on solid surfaces could accelerate the design of formulations with improved sustainability.

**Acknowledgements:** The authors thank Cristobal Galder and Nikolay Christov from Solvay for fruitfull discussions and advice. ANR (Agence Nationale de la Recherche) and CGI (Commissariat à l'Investissement d'Avenir) are gratefully acknowledged for their financial support of this work through Labex SEAM (Science and Engineering for Advanced Materials and devices) ANR 11 LABX 086, ANR 11 IDEX 05 02. This research was supported in part by the Agence Nationale de la Recherche under the contract ANR-15-CE18-0024-01 (ICONS), ANR-17-CE09-0017 (AlveolusMimics) and by Solvay.

## Graphical Abstract

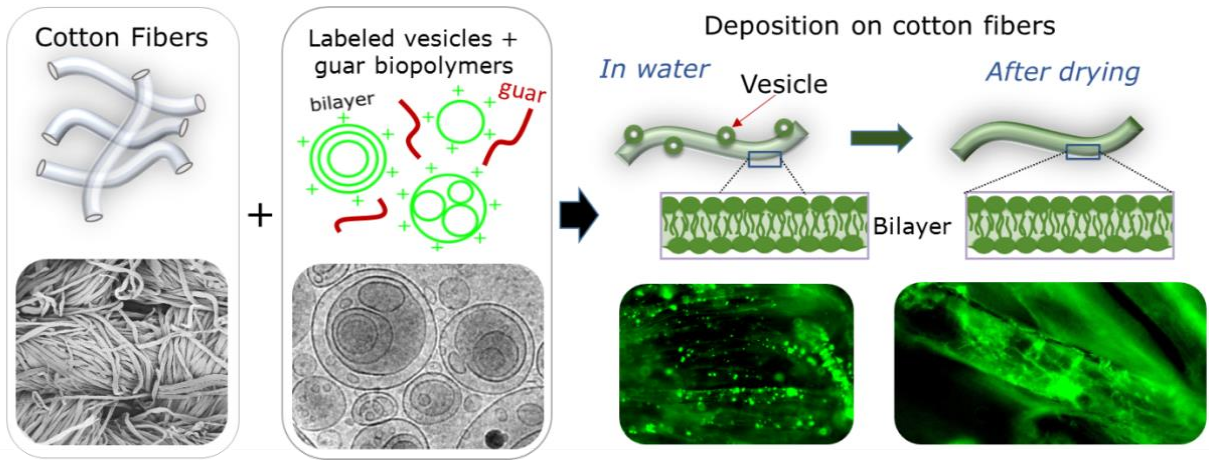